\documentclass[nofootinbib, reprint, amsmath, amssymb, aps, prl, showpacs]{revtex4-1}

\usepackage{graphicx}   % Include figure files
\usepackage{dcolumn}    % Align table columns on decimal point
\usepackage{bm}     % bold math
\usepackage[caption=false]{subfig}

\usepackage{dsfont}     % \mathds for identity matrix

\newcommand\der[2]{\frac{\text{d}{#1}}{\text{d}{#2}}}
\newcommand\pder[2]{\frac{\partial{#1}}{\partial{#2}}}
\newcommand\inv[1]{\frac{1}{#1}}

\newcommand\mx[1]{\bm{#1}}
\newcommand{\id}{\mathds 1}

\newcommand\trans{^T}
\newcommand\Tr{\text{Tr}}

\newcommand\e{\text{e}}
\newcommand\ie{\textit{i.e.}~}
\newcommand\eg{\textit{e.g.}~}
\DeclareRobustCommand{\Chi}{{\mathpalette\irchi\relax}}
\newcommand{\irchi}[2]{\raisebox{\depth}{$#1\chi$}} % inner command, used by \Chi

\newcommand\eq[1]{Eq.~(\ref{eq:#1})}
\newcommand\Eq[1]{Equation~(\ref{eq:#1})}
\newcommand\fig[1]{Fig.~\ref{fig:#1}}
\newcommand\Fig[1]{Figure~\ref{fig:#1}}
\newcommand\rx[1]{(\ref{rx:#1})}

\newcommand{\norm}[1]{\left\lVert#1\right\rVert}
\newcommand{\mean}[1]{\left\langle#1\right\rangle}

\newcounter{defcounter}
\newenvironment{smequation}{%
\addtocounter{equation}{-1}
\refstepcounter{defcounter}

\begin{equation}}
{\end{equation}}

\begin{document}
\title{Giant amplification of noise in fluctuation-induced pattern formation}

\author{Tommaso Biancalani}
\thanks{Present address: Physics of Living Systems, Department of Physics,
Massachusetts Institute of Technology, Cambridge, MA}
\author{Farshid Jafarpour}
\thanks{T. Biancalani and F. Jafarpour contributed equally to this work.}
\author{Nigel Goldenfeld}
\affiliation{Department of Physics, University of Illinois at Urbana-Champaign,
Loomis Laboratory of Physics, 1110 West Green Street, Urbana, Illinois, 61801-3080.}
\affiliation{Carl R. Woese Institute for Genomic Biology, University of Illinois at Urbana-Champaign,
1206 West Gregory Drive, Urbana, Illinois 61801.}
\date{\today}

\begin{abstract}
The amplitude of fluctuation-induced patterns might be expected to be
proportional to the strength of the driving noise, suggesting that such
patterns would be difficult to observe in nature. Here, we show that a
large class of spatially-extended dynamical systems driven by intrinsic
noise can exhibit giant amplification, yielding patterns whose
amplitude is comparable to that of deterministic Turing instabilities.
The giant amplification results from the interplay between noise and
non-orthogonal eigenvectors of the linear stability matrix, yielding
transients that grow with time, and which, when driven by the
ever-present intrinsic noise, lead to persistent large amplitude
patterns. This mechanism provides a robust basis for
fluctuation-induced biological pattern formation based on the Turing
mechanism, without requiring fine tuning of diffusion constants.
\end{abstract}

% \pacs{87.10.Mn, 87.23.Cc, 02.50.-r}
\maketitle

Since the seminal paper of Turing~\cite{turing1952chemical}, it has
been recognized that pattern forming dynamical instabilities could
potentially underlie various examples of biological pattern formation
and development~\cite{koch1994biological}. The Turing mechanism has two
major assumptions: first, that two chemical species behave as an
activator-inhibitor system (but see a recent
extension~\cite{werner2015scaling}), and secondly, that the spatial
diffusion constant of the inhibitor is greater than that of the
activator, typically by two orders of magnitude or
more~\cite{murray2001mathematical}. However, this second condition is
not generally present in experimental
observations~\cite{castets1990experimental, ouyang1991transition}. The
widely-held conclusion is that biological patterns reflect gene
expression and the interplay of developmental processes, so that the
Turing mechanism itself is not generally
operative~\cite{maini2012turing}.

This conclusion relies upon a third assumption of Turing patterns: that
they are deterministic.  However, many biological systems exhibit
strong fluctuations due to demographic stochasticity, arising from
(e.g.) finite population size (ecology) or copy number (gene
expression), and these fluctuations could potentially couple to the
underlying pattern-forming instabilities.  Detailed analysis shows that
the length scale of fluctuation-induced patterns is set by the same
condition as in the deterministic Turing analysis, but remarkably the
pattern exists over a wide range of parameter values, even where the
diffusion constants of activator and inhibitor are of similar
magnitudes \cite{butler2009robust, biancalani2010stochastic,
datta2011stability, ridolfi_noise-induced_2011,
bonachela2012patchiness, butler2012evolutionary}.  These
fluctuation-induced or stochastic patterns arise physically because,
even though the uniform unpatterned state is linearly stable, the
demographic fluctuations are constantly pushing the system slightly
away from its stable fixed point; if the resulting small amplitude
dynamics is dominated by an eigenvalues with a non-zero wavelength, then a spatial pattern can arise.

This mechanism suggests that the amplitude of fluctuation-induced
patterns would be set by $\Omega^{-1/2}$, where $\Omega$ indicates the
population size within a correlation volume of the system, ie. the
spatial patch within which the system can be considered to be well
mixed~\cite{butler2009robust, biancalani2010stochastic}. Thus in
situations where $\Omega \gg 1$, fluctuation-induced patterns might
have a very small amplitude compared to deterministic Turing patterns,
potentially diminishing their relevance for biological and ecological
pattern formation.

\begin{figure}[t]
    \includegraphics[width=.48\textwidth]{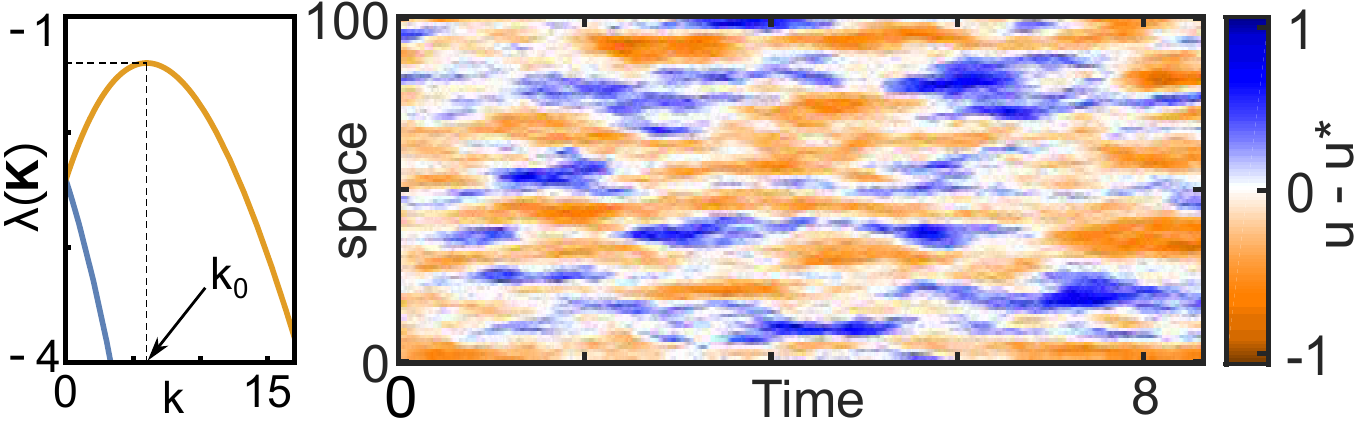}
    \caption{\textbf{(Color online) Turing-like pattern with large
    amplitude and comparable diffusivities}. (right panel) Stochastic
    simulations~\cite{gillespie2013perspective} of a two-species
    model~\rx{main} with diffusivities $\delta_U = 3.9$, $\delta_V =
    3.4\,\delta_U$ and system size $\Omega = 10^4$. Patterns are
    noise-induced as they arise from a stable homogeneous state $u^*$,
    \textit{i.e.}, the eigenvalues $\lambda$ plotted against the
    wavelength $k$ are negative (left panel). However, the pattern
    amplitude results of the order of one (right bar). Other
    parameters: $a = 3$, $b = 5.8$, $c = e = 1$.}
    \label{fig:1}
\end{figure}

The purpose of this Letter is to show that fluctuation-induced Turing
patterns can readily be observed, even when the noise is very small and
the ratio of diffusion constants is close to one.  Specifically we
present an analytical theory showing the presence of giant
amplification, due to an interplay between a separation of time scales
and non-normality of the eigenvectors in the linear stability analysis
about a uniform stable steady state. We present a measure of
non-normality for a general stochastic dynamical system near a stable
fixed point, with a clear geometrical interpretation. We then show that
giant amplification occurs in a wide class of fluctuation-induced
pattern-forming systems.
An example of our key result described below is shown in
Fig.~\ref{fig:1}: stochastic simulations of the generic pattern-forming
model of Ridolfi et al.~\cite{ridolfi2011transient}, performed on a
linear chain of $10^2$ spatial cells, each cell with a system size of
$\Omega=10^4$. Patterns are noise-induced as they arise from a stable
homogeneous state (left panel), but despite the factor $\Omega^{-1/2} =
10^{-2}$ the resulting amplitude is of order unity.

\begin{figure}[t]
    % Graphs are wrong - redo them!
    \includegraphics[width=.48\textwidth]{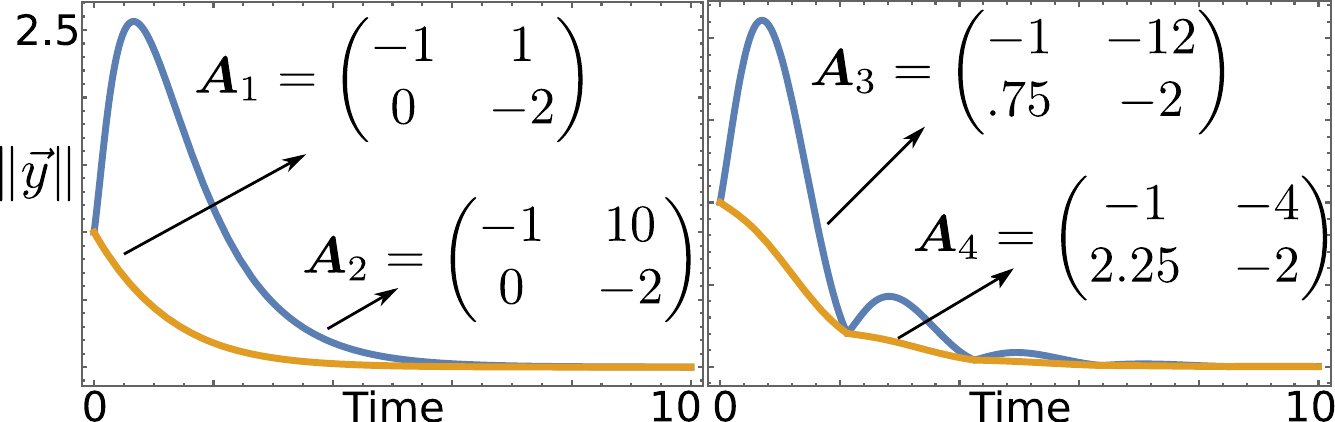}
    \caption{\textbf{(Color online) Stable linear systems can amplify
    perturbations~\cite{neubert1997alternatives}}. Dynamics of the
    Euclidean norm $\norm{\vec y}$ obtained by solving $\dot{\vec y} =
    \mx A_i \vec y$. Reactive systems exhibit transient amplification
    before relaxing to fixed point (blue lines), in contrast with
    conventional response of stable systems (yellow lines). Matrices
    $\mx A_1$ and $\mx A_2$ (respectively $\mx A_3$ and $\mx A_4$) have
    same real (respectively complex conjugate) eigenvalues.}
    \label{fig:2}
\end{figure}

This giant amplification is due to the counterintuitive fact that the
dynamics following a small displacement from a stable fixed point need
not relax back to the fixed point monotonically: there can be an
initial transient amplification if the linear stability matrix is
non-normal: that is, it does not admit an orthogonal set of
eigenvectors (Fig.~\ref{fig:2}). Non-normality has been thoroughly
investigated, at a deterministic level, in fluid
dynamics~\cite{trefethen1993hydrodynamic, trefethen2005spectra}, and in
ecology~\cite{neubert1997alternatives, tang2014reactivity}, and is a
common feature of pattern-forming systems~\cite{neubert2002transient,
ridolfi2011transient}. Low-dimensional stochastic non-normal systems
may also exhibit strong amplification of
noise~\cite{farrell1994variance}. The specific contribution of the
present paper is to systematically analyze the role of non-normality in
fluctuation-induced spatial patterns, and to show that its widespread
occurrence suggests a new way in which fluctuation-induced Turing
patterns may play a wider role in biological and ecological pattern
formation than previously recognized.

\smallskip
\noindent {\it Non-normality in stochastic dynamics:-}
We begin by introducing a measure to quantify the degree of
amplification in a well-mixed stochastic system. Consider the linear
stochastic differential equation for an $m$-component state vector
$\vec y$:
\begin{equation} \label{eq:linsde1}
    \dot{\vec y} = \mx A\, \vec y + \sigma\, \vec \eta(t),
\end{equation}
where the components of $\vec \eta$, are normalized Gaussian white
noises and the model-dependent matrix $\mx A$ has negative real
eigenvalues, $\lambda_i$ ($i = 1, \dots, m$). Therefore, the fixed
point $\vec y_0 = 0$ is stable. The coefficient $\sigma$ represents the
strength of the fluctuations and scales with the system size
$\Omega^{-1/2}$ in the case of demographic noise. \Eq{linsde1} is the
prototypical linearization of stochastic dynamics near a stable fixed
point, and we analyze the mean square displacement from the fixed
point, $\left\langle \norm{\vec y}^2\right\rangle$, where $\norm{\vec
y} = \sqrt{\vec y\trans\vec y}$, is the Euclidean norm.

Since all the eigenvalues of $\mx A$ are negative, under the
deterministic part of \eq{linsde1}, all the components of $\vec y$
decay exponentially to zero along the eigenvectors of $\mx A$, with
decay time scales $\tau_i = \lambda_i^{-1}$. In contrast, the noise
term provides stochastic agitation with a strength proportional to
$\sigma$. One might intuitively expect that an upper bound for
$\left\langle \norm{\vec y}^2\right\rangle$ could be found by replacing
all the eigenvalues by the eigenvalues corresponding to the slowest
decaying mode, $\lambda = \text{max}\{\lambda_i\}$.
Therefore, the norm of $\vec y_u$ with the dynamics $\dot{\vec y}_u =
\lambda\, \vec y_u + \sigma\, \vec \eta(t)$, should provide an upper
bound for $\norm{\vec y}$. The mean square norm of $\vec y_u$ is given
by ($\tau = -\lambda^{-1}$):
\begin{equation}\label{eq:nonreactive}
    \left\langle \norm{\vec y_u}^2\right\rangle = \left\langle \norm{\int_0^{\tau/2}\vec\eta(t)dt}^2\right\rangle = \frac{m}{2}\tau \sigma^2.
\end{equation}

However, this upper bound is only valid when the matrix $\mx A$ is
normal, \ie it has an orthogonal set of eigenvectors (for instance,
Hermitian matrices are normal)~\cite{farrell1994variance}. This can be
understood by analyzing the behavior of \eq{linsde1} in the
deterministic limit ($\sigma = 0$). Although the asymptotic decay rate
of $\norm{\vec y}$ is set by the eigenvalues of $\mx A$, the
instantaneous response is given by the eigenvalues of $\mx H = (\mx A +
\mx A^T)/2$, the Hermitian part of $\mx
A$~\cite{neubert1997alternatives}. If $\mx A$ is non-normal, then the
short-time dynamics of $\norm{\vec y}$ cannot be predicted by the
eigenvalues of $\mx A$. Remarkably, $\mx H$ can admit positive
eigenvalues even though $\mx A$ possesses all negative eigenvalues, in
which case $\norm{\vec y}$ can experience a transient growth, for
suitable initial conditions, before it starts decaying
(Fig.~\ref{fig:2}). This mechanism, sometimes termed as
reactivity~\cite{neubert1997alternatives}, occurs because the
transformation that takes $\vec y$ to the eigenbasis of $\mx A$ is not
unitary if the eigenvectors of $\mx A$ are not orthogonal, and thus does
not preserve the norm of $\vec y$. Clearly, if the stable matrix
amplifies perturbations, the bound~\eqref{eq:nonreactive} cannot hold.

In the presence of noise, this transient effect in the deterministic part
of \eq{linsde1} has a lasting effect on the steady state amplitude of
the stochastic dynamics. This can be demonstrated by solving the steady
state probability density of $\vec y$ for \eq{linsde1}. The detailed
derivation of what follows is presented in the supplemental material
(SM). For every stable matrix $\mx A$, we define a matrix $\mx G$ such
that the Hermitian part of its inverse is the identity, and its product
with $\mx A$ is Hermitian, that is,
\begin{equation}
\begin{split}
    &\inv 2\left(\mx G^{-1}+\left(\mx G^{-1}\right)^T\right) = \id, \quad \left(\mx G\mx A\right)^T = \mx G \mx A.
\end{split}
\end{equation}
Note that $\mx G$ is the identity matrix if $\mx A$ is Hermitian. In
terms of this matrix $\mx G$, the steady state probability density of
$\vec y$ is given by
\begin{equation}
    P\left(\vec y\right) = \sqrt{\det\left(-\frac{\mx G \mx A}{\pi \sigma^2}\right)}\exp\left(\frac{\vec y^{\,T}\mx G\mx A \vec y}{\sigma^2}\right),
\end{equation}
hence the mean square value of $\norm{\vec y}$ is (tr stands for the trace function)
\begin{equation}\label{eq:msv2}
    \mean{\norm{\vec y}^2} = -\frac{\sigma^2}{2} \mathcal{H}(\mx A)\,\text{tr}\left(\mx A^{-1}\right),
\end{equation}
where we have defined the non-normality index $\mathcal{H}$ by:
\begin{equation}\label{eq:index}
    \mathcal{H}(\mx A) =\text{tr}\left(\mx G^{-1} \mx A^{-1}\right)/\text{tr}\left( \mx A^{-1}\right).
\end{equation}

Note that we always have $\mathcal{H}\geq 1$, and $\mathcal H$ is equal
to one if and only if the matrix $\mx A$ is normal. Moreover, the
further $\mx A$ is from normal, the larger is the index $\mathcal{H}$.
In the case of a two-dimensional matrix $\mx A$, the non-normality
index $\mathcal{H}$ simplifies to the following simple expression,
where $\cot \theta$ is the cotangent of the angle between the two
eigenvectors:
\begin{equation}
    \mathcal{H} = 1 + \cot^2(\theta) \left( \frac{\lambda_1 - \lambda_2}{\lambda_1+\lambda_2}\right)^2.
\end{equation}

\begin{figure}
    \centering
    \includegraphics[width=.48\textwidth]{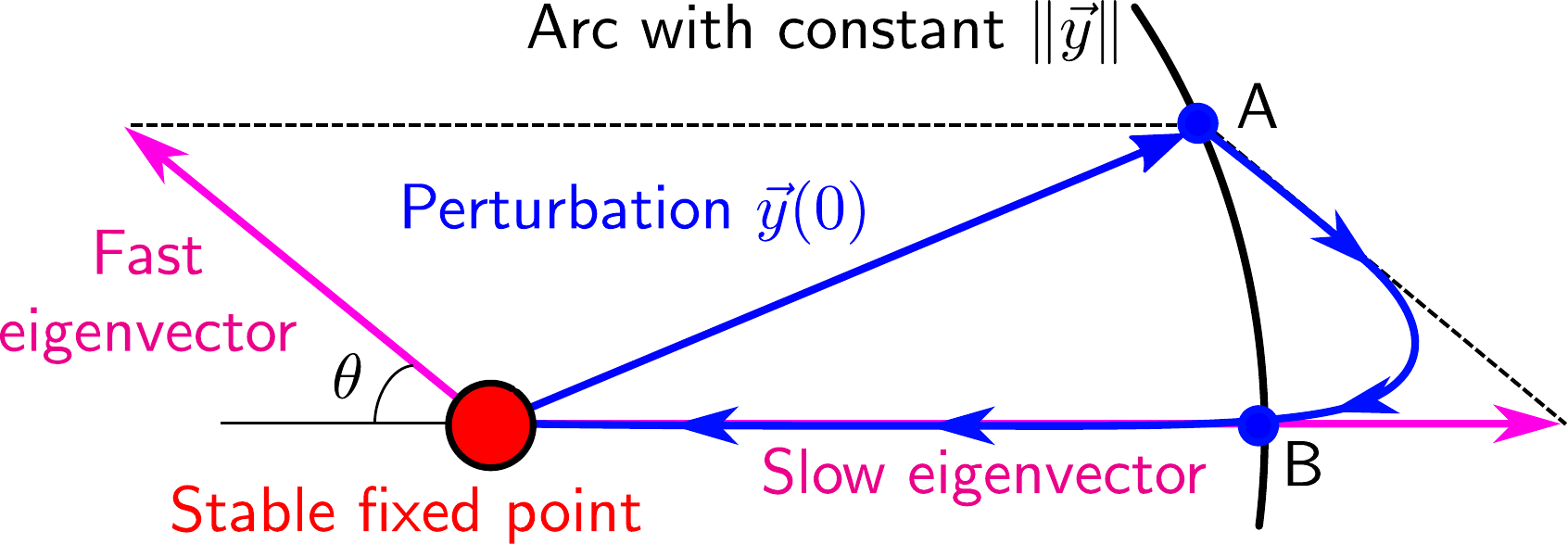}
    \caption{\textbf{(Color online) Transient amplification is caused by
    non-orthogonal eigenvectors and a separation of timescales}. The
    stable fixed point is subject to the perturbation $\vec y(0)$.
    Because of the separation of timescales, the deterministic
    trajectory (blue arrowed line) is initially parallel to the fast
    eigenvector before relaxing to the slow manifold. From $A$ to $B$,
    the trajectory has magnitude greater than $||\vec y_0 ||$.}
    \label{fig:3}
\end{figure}

This expression gives us quantitative understanding about how transient
amplification occurs (\fig{3}). Two ingredients are necessary:
non-orthogonal eigenvectors and a separation of time scales given by
eigenvalues of different magnitudes. If the system is not subject to
noise, suitable initial conditions are also required (\eg the blue
vector in \fig{3}). Because of the separation of time scales, the
component of $\vec y$ along the eigenvector associated with the faster
eigenvalue decays quickly, whereas in the slow direction the dynamics
is approximately constant. However, because of non-orthogonality, the
norm of $\vec y$ instantaneously increases as $\vec y$ moves along the
fast eigenvector, until the slow manifold starts attracting the
trajectory back to fixed point.

%%%%%%%%%%%%%%%%%%%%%%%%%%%%%%%%%%%%%%%%%%%%%%%%%%
\medskip
\noindent {\it Non-normality in spatially-extended pattern formation:-}
We now analyze spatially-extended, diffusively-coupled pattern-forming
systems driven by noise. Specifically, we consider the generic equation
\begin{equation}\label{eq:nonlinear}
    \pder{\vec q}{t} = \vec f(\vec q)+\mx D\nabla^2\vec q+\sigma\vec\xi(\vec x,t),
\end{equation}
where $\vec x$ is a space variable, the vector $\vec q = (q_1, q_2)$,
the diffusion matrix $\mx D = \text{diag}(D_1, D_2)$, and
$\xi_i$\rq{}s, the components of $\vec\xi(\vec x,t)$ are normalized
$\delta$-correlated Gaussian white noises. Also, we assume that $\vec
f(\vec q)$ has a stable fixed point $\vec q^{\,*}$, and all of the
eigenvalues of the linear stability or Jacobian matrix $\mx J = \left.\nabla_{\vec q}
f(\vec q)\right|_{\vec q^{\,*}}$ have negative real part.

Our goal is to show that in the presence of noise,
system~\eqref{eq:nonlinear} exhibits patterns in a parameter regime
where the fixed point $\vec q^{\,*}$ is stable. The stability of $\vec
q^*$ can be inspected by defining the deviation $\vec p = \vec q - \vec
q^{\,*}$ and linearizing near $\vec q\,^*$, yielding
\begin{equation}\label{eq:linear}
    \pder{\vec p}{t} = \mx J\vec p+\mx D\nabla^2\vec p+\sigma\vec\xi(\vec x,t).
\end{equation}
The spatial degrees of freedom can be diagonalized by a Fourier
transform ($\vec x \mapsto \vec k$), resulting in
\begin{equation}\label{eq:fourier}
    \der{\vec p_{\vec k}}{t} = \mx K\vec p_{\vec k}+\sigma\vec\xi(\vec k,t),\qquad \mx K = \mx J - k^2\mx D.
\end{equation}
The equations are now decoupled and are therefore tantamount to
Eq.~\eqref{eq:linsde1}.

We start by reviewing the stability of the deterministic part of
\eq{linear}. If $D_1=D_2$, matrix $\mx D$ is a multiple of the
identity, and the eigenvalues of $\mx K$ will be the eigenvalues of
$\mx J$ shifted by $-k^2D$ for each $\vec k$, resulting in a more
stable operator. However, in the case that the diffusion rates are
sufficiently different, the largest eigenvalue of $\mx K$ can have a
non-monotonic behavior as a function of $\vec k$, and in some cases
have positive eigenvalues for a small range of $\vec k$ peaked around
some non-zero value $\vec k_0$. In this case, the modes near $\vec k_0$ will
grow leading to the formation of deterministic Turing patterns
\cite{turing1952chemical}. Therefore, the formation of deterministic Turing
patterns is dependent on a large separation of the diffusion
constants~\cite{murray2001mathematical,castets1990experimental,
ouyang1991transition}.

In contrast, consider an intermediate scenario with diffusion constants
different enough so that they can cause a non-monotonic behavior for
the largest eigenvalue of $\mx K$ as a function of $\vec k$ peaked
around some value $\vec k_0$, but not enough for the largest eigenvalue
to become positive at any $\vec k$ (left panel of Fig.~\ref{fig:1}). In this
case, all the $\vec k$ modes decay quickly to zero, but the modes with
$\vec k \sim \vec k_0$ decay slower than the others, causing a
transient pattern. In the presence of the noise term $\vec \xi(\vec
k,t)$ in \eq{fourier}, while the modes with smaller eigenvalues decay
quickly to zero, the slow modes drift away from the fixed point under
the influence of the noise. The drift of the $\vec k$ modes near $\vec
k_0$ produces persistent steady-state fluctuation-induced patterns with well-defined
length-scales \cite{butler2009robust, biancalani2010stochastic}. While the stochastic
Turing patterns have a less stringent requirement than the
deterministic Turing patterns for the ratio of the diffusion constants,
their amplitude is limited to the amplitude of the drift under the
noise suppressed by the slow deterministic decay. As discussed in the
previous section, the mean square amplitude is of order $\lambda^{-1}
\sigma^2$, unless we can show that the system is non-normal.

We now prove that in order for a system described by \eq{nonlinear} to
produce stochastic patterns, it is necessary for the matrix $\mx J$ in
\eq{linear} to be non-normal. We show this by finding a lower bound on
the difference between the largest eigenvalue of $\mx H = (\mx J + \mx
J^{T})/2$ and that of matrix $\mx J$. The proof relies on the fact that
for the system to exhibit stochastic patterns, the real part of the
largest eigenvalue, $\lambda_1$, of $\mx K$ as a function of the wave
vector $\vec k$ should peak at some value $\vec k_0 \neq
0$~\cite{mckane2014stochastic, biancalani2010stochastic}, and
therefore, $\delta = \Re(\lambda_1(\mx K_0)) - \Re(\lambda_1(\mx J)) >
0$, for $\mx K_0 = \mx K(\vec k_0)$. It is a well known fact that the
real part of the largest eigenvalue of a matrix is less than or equal
to that of its Hermitian part (\eg see Ref.~\cite{adam2006eigenvalue}),
therefore, $\Re(\lambda_1(\mx K_0))\leq \lambda_1(\mx H -k_0^2 \mx D)$.
Since both $\mx H$ and $ -k_0^2 \mx D$ are Hermitian, by Weyl
inequality $\lambda_1(\mx H -k_0^2 \mx D) \leq \lambda_1(\mx H)
+\lambda_1(-k_0^2 \mx D) = \lambda_1(\mx H) - k_0^2 D_{min}$. Adding
$\,k_0^2\, D_{min} - \Re(\lambda_1(\mx J))$ to both sides of this
inequality, we arrive at
\begin{equation}\label{nonnormality_bound}
    \lambda_1(\mx H) - \Re(\lambda_1(\mx J)) \geq \delta +  k_0^2 D_{min}.
\end{equation}
Since the non-normality of $\mx J$ should be independent of the
diffusion constants, this lower bound can be extended to the supremum
of the right hand side of the inequality \eqref{nonnormality_bound}
over all the matrices $\mx D$ that produce spatial patterns and their
corresponding $\vec k_0$. In particular, if a system admits
deterministic Turing patterns for some set of diffusion constants, \ie
$\Re(\lambda_1(\mx K_0)) >0$, $\delta$ would be greater than $ -
\Re(\lambda_1(\mx J))$, and therefore $\mx J$ would be reactive (this
special case was previously proven by Neubert et
al.~\cite{neubert2002transient}). In this case, if experimentally
measured values of diffusion constants do not fall within the Turing
pattern regime, the system is still reactive and capable of exhibiting
amplified stochastic patterns.

%%%%%%%%%%%%%%%%%%%%%%%%%%%%%%%%%%%%%%%%%%%%%%%
\smallskip
\noindent {\it Stochastic extension of model by Ridolfi et al.\,:-}
Finally, we apply our theory to a concrete model that is representative
of a large class of systems. On a deterministic level, the model is
given by Eq.~\eqref{eq:nonlinear} with two species $U$ and $V$ with
densities $\vec q = (u,\,v)$, and $\vec f(u, v) = \left(u (a u v -
e),\, v(b - c u^2 v)\right)$, with $a,b,c, e >
0$~\cite{ridolfi2011transient}. The corresponding stochastic model is
defined by considering the following individual-level processes that
occur on a discretized $D$-dimensional space with $L^D$ lattice sites,
\begin{figure}[t]
    \includegraphics[width=.48\textwidth]{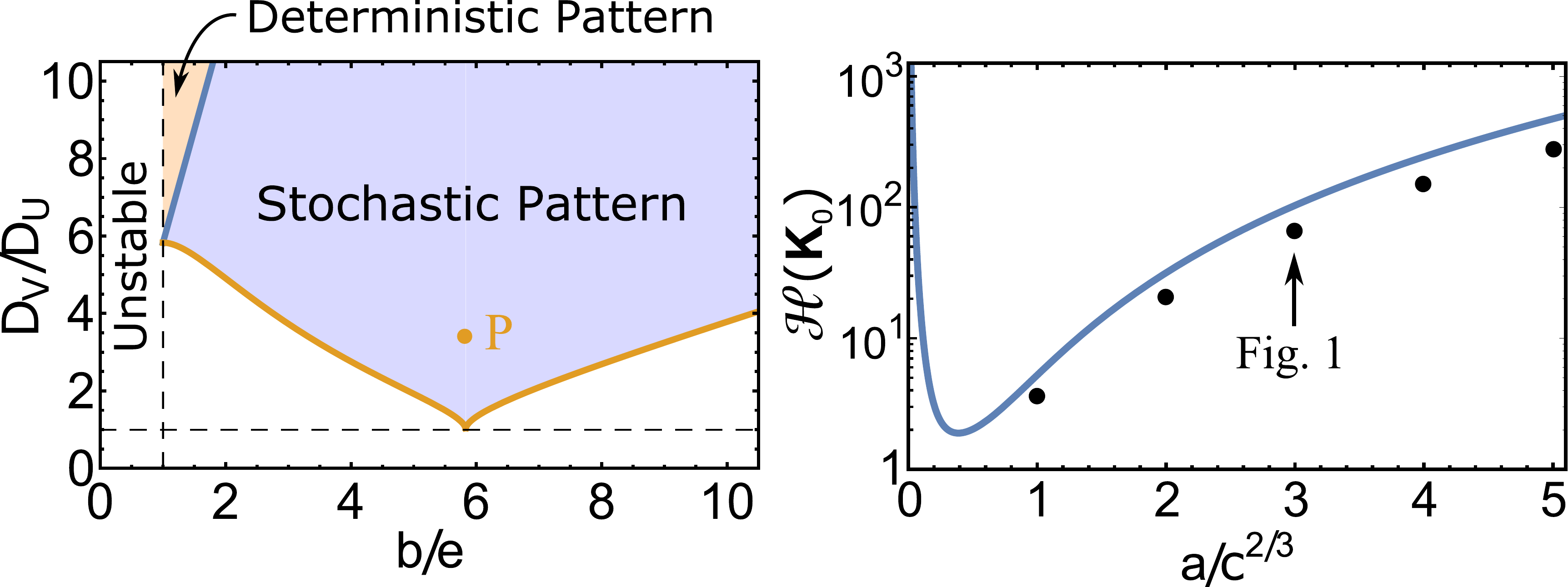}
    \caption{\textbf{(Color online) Stochasticity allows pattern formation for similar diffusivities}. (left) Phase diagram of model \rx{main} showing that the pattern forming behavior of this model depends only on the ratios $b/a$ and $D_V/D_U$ (see SM for analytic expression for the boundaries). (right) Semi-log plot of non-normality index for the point $P$ as a function of $a/c^{2/3}$. Black markers are amplifications measured in simulation.}
    \label{fig:4}
\end{figure}
\begin{equation}\label{rx:main}
\begin{split}
    &2U_i+V_i\xrightarrow{a}3U_i+V_i,\hspace{0.05\textwidth}
    V_i\xrightarrow{b}2V_i,\\
    &U_i\xrightarrow{e}\varnothing,\hspace{0.05\textwidth}
    2V_i+2U_i\xrightarrow{c}V_i+2U_i,\\
    &U_i\xrightarrow{\delta_U}U_j,\hspace{0.05\textwidth}
    V_i\xrightarrow{\delta_V}V_j,\hspace{0.05\textwidth} j\in\langle i\rangle
\end{split}
\end{equation}
where $U_i$ and $V_i$ are the species $U$ and $V$ on the site $i$ for
$i = 1\dots L^D$ and $\langle i\rangle$ is the set of sites neighboring
$i$. The state of the system is specified by the concentration vectors
$\vec q_i\equiv (u_i, v_i) \equiv (U_i, V_i)/\Omega$, where $\Omega$ is
the volume of each site. The diffusion rates $\delta_u$ and $\delta_v$
are related to the diffusion constants by $(\delta_u,\,\delta_v) =
(D_U, D_V)/\Omega^{2/D}$. The discrete-space version of Eqs.
\eqref{eq:nonlinear}, \eqref{eq:linear} and \eqref{eq:fourier} are
derived by expanding in powers of $\Omega^{-1/2}$ the master equation
corresponding to scheme~\eqref{rx:main} (see the SM for the
derivations).

The pattern forming behavior of the model described by \rx{main} only
depends on the ratio of the diffusion constants $D_V/D_U$ and the ratio
of the reaction rates of the two linear reactions $b/e$. The left panel of \Fig{4}
shows the regime of parameters in which the system exhibits either
stochastic or deterministic Turing patterns. As expected, deterministic
patterns emerge only when the ratio $D_V/D_U$ of diffusion constants is
very large (above the blue line in \fig{4} which steeply grows outside
of the figure), while the requirement on this ratio for the stochastic
patterns is drastically reduced (see the SM for analytic expressions
for the boundaries). In the absence of the non-normality effect, one
would expect that only stochastic patterns with parameters very
close to the deterministic regime would be observed, since far from
this regime, the amplitude of the patterns would be too small to
detect.

However, since for all $b/e>1$, there is a $D_V/D_U$ above which the
system exhibits deterministic Turing patterns, $\mx J$ is reactive.
Therefore, even when the system is far from the parameter regime of
deterministic patterns, the amplitude of the stochastic patterns is far
larger than what one would expect from the analysis of the eigenvalues
from \eq{nonreactive}. We can see this by analyzing the amplitude of
the patterns at the point $P$ in \fig{4}. This point has ratios
$b/e = 5.8$ and $D_V/D_U = 3.4$ and is chosen to be very far from the
deterministic Turing pattern regime. At this $b/e$ ratio, the ratio of
the diffusion constants has to be at least ten times larger than the
chosen value for the system to exhibit deterministic Turing patterns.
The amplitude of the patterns as determined by \eq{msv2} is dependent
on the eigenvalues of $\mx K$ (fixed by the choice of the point $P$)
and the non-normality index $\mathcal{H}(\mx K)$ which can be tuned by
changing the ratio $a/c^{2/3}$ without changing the point $P$ (see SM
for the analytic expression). The right panel of \Fig{4} shows that the
amplification of stochastic patterns for the point $P$ varies over
orders of magnitude for a small range of $a/c^{2/3}$.

The right panel of \Fig{2} shows the time series of the amplified stochastic Turing
patterns in the concentration of the species $U$, in a simulation of
our model in one dimension. The mean square amplitude of these spatial
patterns is about $0.21$, while the upper bound for the amplitude of
the pattern in the absence of reactivity from \eq{nonreactive} is
$2.5\times10^{-3}$. The non-normality index $\mathcal{H}$ of the
slowest Fourier mode $k_0 = 6$ is about $103$ justifying the two order
of magnitude amplification in the amplitude of the stochastic patterns
(see the right panel of \fig{4}).\\

%%%%%%%%%%%%%%%%%%%%%%%%%%%%%%%%%%%%%%%%%%%%%%%

In conclusion, fluctuation-induced Turing patterns have larger amplitude than previously expected, even when the ratio of the diffusion coefficients is far from the requirement for deterministic Turing patterns. This large amplitude is due to non-normality of the type of interactions that are required for a system to produce Turing-like patterns. We have introduced a new measure of non-normality that is applicable to all stochastic dynamical systems and measures the amplification of the expected value of the distance that a non-equilibrium system maintains from its fixed point at steady state. We have used this measure to quantify the effect of non-normality on stochastic Turing patterns and explain the unexpectedly large amplitude observed in the simulations. By analyzing an example of an activator-inhibitor system, we have shown that the demographic stochasticity drastically expands the range of parameters in which the system exhibits Turing-like patterns, and that these patterns have amplitudes that are orders of magnitude larger that expected in all but a narrow region in parameter space. We conclude that fluctuation-induced Turing patterns can readily be observed, and therefore, provide a potential mechanism explaining a wide range of patterns formations observed in ecology, biology, and development

% https://journals.aps.org/authors/byline-addresses-footnotes-acknowledgments-statements-about-authors-h22
This work was supported by the National Aeronautics and Space
Administration Astrobiology Institute (NAI) under Cooperative Agreement
No. NNA13AA91A issued through the Science Mission Directorate.  T.B
acknowledges partial funding from the National Science Foundation under
Grant No. PHY-105515. T. B. and F. J. Contributed equally to this work.
\bibliographystyle{apsrev4-1}
\bibliography{literature}

\begin{thebibliography}{25}%
\makeatletter
\providecommand \@ifxundefined [1]{%
 \@ifx{#1\undefined}
}%
\providecommand \@ifnum [1]{%
 \ifnum #1\expandafter \@firstoftwo
 \else \expandafter \@secondoftwo
 \fi
}%
\providecommand \@ifx [1]{%
 \ifx #1\expandafter \@firstoftwo
 \else \expandafter \@secondoftwo
 \fi
}%
\providecommand \natexlab [1]{#1}%
\providecommand \enquote  [1]{``#1''}%
\providecommand \bibnamefont  [1]{#1}%
\providecommand \bibfnamefont [1]{#1}%
\providecommand \citenamefont [1]{#1}%
\providecommand \href@noop [0]{\@secondoftwo}%
\providecommand \href [0]{\begingroup \@sanitize@url \@href}%
\providecommand \@href[1]{\@@startlink{#1}\@@href}%
\providecommand \@@href[1]{\endgroup#1\@@endlink}%
\providecommand \@sanitize@url [0]{\catcode `\\12\catcode `\$12\catcode
  `\&12\catcode `\#12\catcode `\^12\catcode `\_12\catcode `\%12\relax}%
\providecommand \@@startlink[1]{}%
\providecommand \@@endlink[0]{}%
\providecommand \url  [0]{\begingroup\@sanitize@url \@url }%
\providecommand \@url [1]{\endgroup\@href {#1}{\urlprefix }}%
\providecommand \urlprefix  [0]{URL }%
\providecommand \Eprint [0]{\href }%
\providecommand \doibase [0]{http://dx.doi.org/}%
\providecommand \selectlanguage [0]{\@gobble}%
\providecommand \bibinfo  [0]{\@secondoftwo}%
\providecommand \bibfield  [0]{\@secondoftwo}%
\providecommand \translation [1]{[#1]}%
\providecommand \BibitemOpen [0]{}%
\providecommand \bibitemStop [0]{}%
\providecommand \bibitemNoStop [0]{.\EOS\space}%
\providecommand \EOS [0]{\spacefactor3000\relax}%
\providecommand \BibitemShut  [1]{\csname bibitem#1\endcsname}%
\let\auto@bib@innerbib\@empty
%</preamble>
\bibitem [{\citenamefont {Turing}(1952)}]{turing1952chemical}%
  \BibitemOpen
  \bibfield  {author} {\bibinfo {author} {\bibfnamefont {A.~M.}\ \bibnamefont
  {Turing}},\ }\href@noop {} {\bibfield  {journal} {\bibinfo  {journal}
  {Philos. Trans. R. Soc. London, Ser. B}\ }\textbf {\bibinfo {volume} {237}},\
  \bibinfo {pages} {37} (\bibinfo {year} {1952})}\BibitemShut {NoStop}%
\bibitem [{\citenamefont {Koch}\ and\ \citenamefont
  {Meinhardt}(1994)}]{koch1994biological}%
  \BibitemOpen
  \bibfield  {author} {\bibinfo {author} {\bibfnamefont {A.}~\bibnamefont
  {Koch}}\ and\ \bibinfo {author} {\bibfnamefont {H.}~\bibnamefont
  {Meinhardt}},\ }\href@noop {} {\bibfield  {journal} {\bibinfo  {journal}
  {Rev. Mod. Phys.}\ }\textbf {\bibinfo {volume} {66}},\ \bibinfo {pages}
  {1481} (\bibinfo {year} {1994})}\BibitemShut {NoStop}%
\bibitem [{\citenamefont {Werner}\ \emph {et~al.}(2015)\citenamefont {Werner},
  \citenamefont {St{\"u}ckemann}, \citenamefont {Amigo}, \citenamefont {Rink},
  \citenamefont {J{\"u}licher},\ and\ \citenamefont
  {Friedrich}}]{werner2015scaling}%
  \BibitemOpen
  \bibfield  {author} {\bibinfo {author} {\bibfnamefont {S.}~\bibnamefont
  {Werner}}, \bibinfo {author} {\bibfnamefont {T.}~\bibnamefont
  {St{\"u}ckemann}}, \bibinfo {author} {\bibfnamefont {M.~B.}\ \bibnamefont
  {Amigo}}, \bibinfo {author} {\bibfnamefont {J.~C.}\ \bibnamefont {Rink}},
  \bibinfo {author} {\bibfnamefont {F.}~\bibnamefont {J{\"u}licher}}, \ and\
  \bibinfo {author} {\bibfnamefont {B.~M.}\ \bibnamefont {Friedrich}},\
  }\href@noop {} {\bibfield  {journal} {\bibinfo  {journal} {Phys. Rev. Lett.}\
  }\textbf {\bibinfo {volume} {114}},\ \bibinfo {pages} {138101} (\bibinfo
  {year} {2015})}\BibitemShut {NoStop}%
\bibitem [{\citenamefont {Murray}(2001)}]{murray2001mathematical}%
  \BibitemOpen
  \bibfield  {author} {\bibinfo {author} {\bibfnamefont {J.~D.}\ \bibnamefont
  {Murray}},\ }\href@noop {} {\emph {\bibinfo {title} {Mathematical Biology. II
  Spatial Models and Biomedical Applications $\{$Interdisciplinary Applied
  Mathematics V. 18$\}$}}}\ (\bibinfo  {publisher} {Springer-Verlag New York
  Incorporated},\ \bibinfo {year} {2001})\BibitemShut {NoStop}%
\bibitem [{\citenamefont {Castets}\ \emph {et~al.}(1990)\citenamefont
  {Castets}, \citenamefont {Dulos}, \citenamefont {Boissonade},\ and\
  \citenamefont {De~Kepper}}]{castets1990experimental}%
  \BibitemOpen
  \bibfield  {author} {\bibinfo {author} {\bibfnamefont {V.}~\bibnamefont
  {Castets}}, \bibinfo {author} {\bibfnamefont {E.}~\bibnamefont {Dulos}},
  \bibinfo {author} {\bibfnamefont {J.}~\bibnamefont {Boissonade}}, \ and\
  \bibinfo {author} {\bibfnamefont {P.}~\bibnamefont {De~Kepper}},\ }\href@noop
  {} {\bibfield  {journal} {\bibinfo  {journal} {Phys. Rev. Lett.}\ }\textbf
  {\bibinfo {volume} {64}},\ \bibinfo {pages} {2953} (\bibinfo {year}
  {1990})}\BibitemShut {NoStop}%
\bibitem [{\citenamefont {Ouyang}\ and\ \citenamefont
  {Swinney}(1991)}]{ouyang1991transition}%
  \BibitemOpen
  \bibfield  {author} {\bibinfo {author} {\bibfnamefont {Q.}~\bibnamefont
  {Ouyang}}\ and\ \bibinfo {author} {\bibfnamefont {H.~L.}\ \bibnamefont
  {Swinney}},\ }\href@noop {} {\bibfield  {journal} {\bibinfo  {journal}
  {Nature}\ }\textbf {\bibinfo {volume} {352}},\ \bibinfo {pages} {610}
  (\bibinfo {year} {1991})}\BibitemShut {NoStop}%
\bibitem [{\citenamefont {Maini}\ \emph {et~al.}(2012)\citenamefont {Maini},
  \citenamefont {Woolley}, \citenamefont {Baker}, \citenamefont {Gaffney},\
  and\ \citenamefont {Lee}}]{maini2012turing}%
  \BibitemOpen
  \bibfield  {author} {\bibinfo {author} {\bibfnamefont {P.~K.}\ \bibnamefont
  {Maini}}, \bibinfo {author} {\bibfnamefont {T.~E.}\ \bibnamefont {Woolley}},
  \bibinfo {author} {\bibfnamefont {R.~E.}\ \bibnamefont {Baker}}, \bibinfo
  {author} {\bibfnamefont {E.~A.}\ \bibnamefont {Gaffney}}, \ and\ \bibinfo
  {author} {\bibfnamefont {S.~S.}\ \bibnamefont {Lee}},\ }\href@noop {}
  {\bibfield  {journal} {\bibinfo  {journal} {Interface focus}\ ,\ \bibinfo
  {pages} {rsfs20110113}} (\bibinfo {year} {2012})}\BibitemShut {NoStop}%
\bibitem [{\citenamefont {Butler}\ and\ \citenamefont
  {Goldenfeld}(2009)}]{butler2009robust}%
  \BibitemOpen
  \bibfield  {author} {\bibinfo {author} {\bibfnamefont {T.}~\bibnamefont
  {Butler}}\ and\ \bibinfo {author} {\bibfnamefont {N.}~\bibnamefont
  {Goldenfeld}},\ }\href@noop {} {\bibfield  {journal} {\bibinfo  {journal}
  {Phys. Rev. E}\ }\textbf {\bibinfo {volume} {80}},\ \bibinfo {pages} {030902}
  (\bibinfo {year} {2009})}\BibitemShut {NoStop}%
\bibitem [{\citenamefont {Biancalani}\ \emph {et~al.}(2010)\citenamefont
  {Biancalani}, \citenamefont {Fanelli},\ and\ \citenamefont
  {Di~Patti}}]{biancalani2010stochastic}%
  \BibitemOpen
  \bibfield  {author} {\bibinfo {author} {\bibfnamefont {T.}~\bibnamefont
  {Biancalani}}, \bibinfo {author} {\bibfnamefont {D.}~\bibnamefont {Fanelli}},
  \ and\ \bibinfo {author} {\bibfnamefont {F.}~\bibnamefont {Di~Patti}},\
  }\href@noop {} {\bibfield  {journal} {\bibinfo  {journal} {Phys. Rev. E}\
  }\textbf {\bibinfo {volume} {81}},\ \bibinfo {pages} {046215} (\bibinfo
  {year} {2010})}\BibitemShut {NoStop}%
\bibitem [{\citenamefont {Datta}\ \emph {et~al.}(2011)\citenamefont {Datta},
  \citenamefont {Delius}, \citenamefont {Law},\ and\ \citenamefont
  {Plank}}]{datta2011stability}%
  \BibitemOpen
  \bibfield  {author} {\bibinfo {author} {\bibfnamefont {S.}~\bibnamefont
  {Datta}}, \bibinfo {author} {\bibfnamefont {G.~W.}\ \bibnamefont {Delius}},
  \bibinfo {author} {\bibfnamefont {R.}~\bibnamefont {Law}}, \ and\ \bibinfo
  {author} {\bibfnamefont {M.~J.}\ \bibnamefont {Plank}},\ }\href@noop {}
  {\bibfield  {journal} {\bibinfo  {journal} {J. Math. Bio.}\ }\textbf
  {\bibinfo {volume} {63}},\ \bibinfo {pages} {779} (\bibinfo {year}
  {2011})}\BibitemShut {NoStop}%
\bibitem [{\citenamefont {Ridolfi}\ \emph
  {et~al.}(2011{\natexlab{a}})\citenamefont {Ridolfi}, \citenamefont
  {D'Odorico},\ and\ \citenamefont {Laio}}]{ridolfi_noise-induced_2011}%
  \BibitemOpen
  \bibfield  {author} {\bibinfo {author} {\bibfnamefont {L.}~\bibnamefont
  {Ridolfi}}, \bibinfo {author} {\bibfnamefont {P.}~\bibnamefont {D'Odorico}},
  \ and\ \bibinfo {author} {\bibfnamefont {F.}~\bibnamefont {Laio}},\
  }\href@noop {} {\emph {\bibinfo {title} {Noise-Induced Phenomena in the
  Environmental Sciences}}}\ (\bibinfo  {publisher} {Cambridge University
  Press},\ \bibinfo {address} {Cambridge},\ \bibinfo {year} {2011})\BibitemShut
  {NoStop}%
\bibitem [{\citenamefont {Bonachela}\ \emph {et~al.}(2012)\citenamefont
  {Bonachela}, \citenamefont {Mu{\~n}oz},\ and\ \citenamefont
  {Levin}}]{bonachela2012patchiness}%
  \BibitemOpen
  \bibfield  {author} {\bibinfo {author} {\bibfnamefont {J.~A.}\ \bibnamefont
  {Bonachela}}, \bibinfo {author} {\bibfnamefont {M.~A.}\ \bibnamefont
  {Mu{\~n}oz}}, \ and\ \bibinfo {author} {\bibfnamefont {S.~A.}\ \bibnamefont
  {Levin}},\ }\href@noop {} {\bibfield  {journal} {\bibinfo  {journal} {J.
  Stat. Phys.}\ }\textbf {\bibinfo {volume} {148}},\ \bibinfo {pages} {724}
  (\bibinfo {year} {2012})}\BibitemShut {NoStop}%
\bibitem [{\citenamefont {Butler}\ \emph {et~al.}(2012)\citenamefont {Butler},
  \citenamefont {Benayoun}, \citenamefont {Wallace}, \citenamefont {van
  Drongelen}, \citenamefont {Goldenfeld},\ and\ \citenamefont
  {Cowan}}]{butler2012evolutionary}%
  \BibitemOpen
  \bibfield  {author} {\bibinfo {author} {\bibfnamefont {T.~C.}\ \bibnamefont
  {Butler}}, \bibinfo {author} {\bibfnamefont {M.}~\bibnamefont {Benayoun}},
  \bibinfo {author} {\bibfnamefont {E.}~\bibnamefont {Wallace}}, \bibinfo
  {author} {\bibfnamefont {W.}~\bibnamefont {van Drongelen}}, \bibinfo {author}
  {\bibfnamefont {N.}~\bibnamefont {Goldenfeld}}, \ and\ \bibinfo {author}
  {\bibfnamefont {J.}~\bibnamefont {Cowan}},\ }\href@noop {} {\bibfield
  {journal} {\bibinfo  {journal} {Proc. Natl. Acad. Sci. USA}\ }\textbf
  {\bibinfo {volume} {109}},\ \bibinfo {pages} {606} (\bibinfo {year}
  {2012})}\BibitemShut {NoStop}%
\bibitem [{\citenamefont {Gillespie}\ \emph {et~al.}(2013)\citenamefont
  {Gillespie}, \citenamefont {Hellander},\ and\ \citenamefont
  {Petzold}}]{gillespie2013perspective}%
  \BibitemOpen
  \bibfield  {author} {\bibinfo {author} {\bibfnamefont {D.~T.}\ \bibnamefont
  {Gillespie}}, \bibinfo {author} {\bibfnamefont {A.}~\bibnamefont
  {Hellander}}, \ and\ \bibinfo {author} {\bibfnamefont {L.~R.}\ \bibnamefont
  {Petzold}},\ }\href@noop {} {\bibfield  {journal} {\bibinfo  {journal} {J.
  Chem. Phys.}\ }\textbf {\bibinfo {volume} {138}},\ \bibinfo {pages} {170901}
  (\bibinfo {year} {2013})}\BibitemShut {NoStop}%
\bibitem [{\citenamefont {Ridolfi}\ \emph
  {et~al.}(2011{\natexlab{b}})\citenamefont {Ridolfi}, \citenamefont
  {Camporeale}, \citenamefont {D'Odorico},\ and\ \citenamefont
  {Laio}}]{ridolfi2011transient}%
  \BibitemOpen
  \bibfield  {author} {\bibinfo {author} {\bibfnamefont {L.}~\bibnamefont
  {Ridolfi}}, \bibinfo {author} {\bibfnamefont {C.}~\bibnamefont {Camporeale}},
  \bibinfo {author} {\bibfnamefont {P.}~\bibnamefont {D'Odorico}}, \ and\
  \bibinfo {author} {\bibfnamefont {F.}~\bibnamefont {Laio}},\ }\href@noop {}
  {\bibfield  {journal} {\bibinfo  {journal} {Eur. Phys. Lett.}\ }\textbf
  {\bibinfo {volume} {95}},\ \bibinfo {pages} {18003} (\bibinfo {year}
  {2011}{\natexlab{b}})}\BibitemShut {NoStop}%
\bibitem [{\citenamefont {Trefethen}\ \emph {et~al.}(1993)\citenamefont
  {Trefethen}, \citenamefont {Trefethen}, \citenamefont {Reddy}, \citenamefont
  {Driscoll} \emph {et~al.}}]{trefethen1993hydrodynamic}%
  \BibitemOpen
  \bibfield  {author} {\bibinfo {author} {\bibfnamefont {L.}~\bibnamefont
  {Trefethen}}, \bibinfo {author} {\bibfnamefont {A.}~\bibnamefont
  {Trefethen}}, \bibinfo {author} {\bibfnamefont {S.}~\bibnamefont {Reddy}},
  \bibinfo {author} {\bibfnamefont {T.}~\bibnamefont {Driscoll}},  \emph
  {et~al.},\ }\href@noop {} {\bibfield  {journal} {\bibinfo  {journal}
  {Science}\ }\textbf {\bibinfo {volume} {261}},\ \bibinfo {pages} {578}
  (\bibinfo {year} {1993})}\BibitemShut {NoStop}%
\bibitem [{\citenamefont {Trefethen}\ and\ \citenamefont
  {Embree}(2005)}]{trefethen2005spectra}%
  \BibitemOpen
  \bibfield  {author} {\bibinfo {author} {\bibfnamefont {L.~N.}\ \bibnamefont
  {Trefethen}}\ and\ \bibinfo {author} {\bibfnamefont {M.}~\bibnamefont
  {Embree}},\ }\href@noop {} {\emph {\bibinfo {title} {Spectra and
  pseudospectra: the behavior of nonnormal matrices and operators}}}\ (\bibinfo
   {publisher} {Princeton University Press},\ \bibinfo {year}
  {2005})\BibitemShut {NoStop}%
\bibitem [{\citenamefont {Neubert}\ and\ \citenamefont
  {Caswell}(1997)}]{neubert1997alternatives}%
  \BibitemOpen
  \bibfield  {author} {\bibinfo {author} {\bibfnamefont {M.~G.}\ \bibnamefont
  {Neubert}}\ and\ \bibinfo {author} {\bibfnamefont {H.}~\bibnamefont
  {Caswell}},\ }\href@noop {} {\bibfield  {journal} {\bibinfo  {journal}
  {Ecology}\ }\textbf {\bibinfo {volume} {78}},\ \bibinfo {pages} {653}
  (\bibinfo {year} {1997})}\BibitemShut {NoStop}%
\bibitem [{\citenamefont {Tang}\ and\ \citenamefont
  {Allesina}(2014)}]{tang2014reactivity}%
  \BibitemOpen
  \bibfield  {author} {\bibinfo {author} {\bibfnamefont {S.}~\bibnamefont
  {Tang}}\ and\ \bibinfo {author} {\bibfnamefont {S.}~\bibnamefont
  {Allesina}},\ }\href@noop {} {\bibfield  {journal} {\bibinfo  {journal}
  {Population Dynamics}\ }\textbf {\bibinfo {volume} {2}},\ \bibinfo {pages}
  {21} (\bibinfo {year} {2014})}\BibitemShut {NoStop}%
\bibitem [{\citenamefont {Neubert}\ \emph {et~al.}(2002)\citenamefont
  {Neubert}, \citenamefont {Caswell},\ and\ \citenamefont
  {Murray}}]{neubert2002transient}%
  \BibitemOpen
  \bibfield  {author} {\bibinfo {author} {\bibfnamefont {M.~G.}\ \bibnamefont
  {Neubert}}, \bibinfo {author} {\bibfnamefont {H.}~\bibnamefont {Caswell}}, \
  and\ \bibinfo {author} {\bibfnamefont {J.}~\bibnamefont {Murray}},\
  }\href@noop {} {\bibfield  {journal} {\bibinfo  {journal} {Mathematical
  biosciences}\ }\textbf {\bibinfo {volume} {175}},\ \bibinfo {pages} {1}
  (\bibinfo {year} {2002})}\BibitemShut {NoStop}%
\bibitem [{\citenamefont {Farrell}\ and\ \citenamefont
  {Ioannou}(1994)}]{farrell1994variance}%
  \BibitemOpen
  \bibfield  {author} {\bibinfo {author} {\bibfnamefont {B.~F.}\ \bibnamefont
  {Farrell}}\ and\ \bibinfo {author} {\bibfnamefont {P.~J.}\ \bibnamefont
  {Ioannou}},\ }\href@noop {} {\bibfield  {journal} {\bibinfo  {journal} {Phys.
  Rev. Lett.}\ }\textbf {\bibinfo {volume} {72}},\ \bibinfo {pages} {1188}
  (\bibinfo {year} {1994})}\BibitemShut {NoStop}%
\bibitem [{\citenamefont {McKane}\ \emph {et~al.}(2014)\citenamefont {McKane},
  \citenamefont {Biancalani},\ and\ \citenamefont
  {Rogers}}]{mckane2014stochastic}%
  \BibitemOpen
  \bibfield  {author} {\bibinfo {author} {\bibfnamefont {A.~J.}\ \bibnamefont
  {McKane}}, \bibinfo {author} {\bibfnamefont {T.}~\bibnamefont {Biancalani}},
  \ and\ \bibinfo {author} {\bibfnamefont {T.}~\bibnamefont {Rogers}},\
  }\href@noop {} {\bibfield  {journal} {\bibinfo  {journal} {Bull. Math.
  Biol.}\ }\textbf {\bibinfo {volume} {76}},\ \bibinfo {pages} {895} (\bibinfo
  {year} {2014})}\BibitemShut {NoStop}%
\bibitem [{\citenamefont {Adam}\ and\ \citenamefont
  {Tsatsomeros}(2006)}]{adam2006eigenvalue}%
  \BibitemOpen
  \bibfield  {author} {\bibinfo {author} {\bibfnamefont {M.}~\bibnamefont
  {Adam}}\ and\ \bibinfo {author} {\bibfnamefont {M.~J.}\ \bibnamefont
  {Tsatsomeros}},\ }\href@noop {} {\bibfield  {journal} {\bibinfo  {journal}
  {Electron. J. Linear Algebra}\ }\textbf {\bibinfo {volume} {15}},\ \bibinfo
  {pages} {239} (\bibinfo {year} {2006})}\BibitemShut {NoStop}%
\bibitem [{\citenamefont {Gardiner}(2009)}]{Gardiner2009}%
  \BibitemOpen
  \bibfield  {author} {\bibinfo {author} {\bibfnamefont {C.~W.}\ \bibnamefont
  {Gardiner}},\ }\href@noop {} {\emph {\bibinfo {title} {{Handbook of
  Stochastic Methods for Physics, Chemistry and the Natural Sciences}}}},\
  \bibinfo {edition} {4th}\ ed.\ (\bibinfo  {publisher} {Springer},\ \bibinfo
  {address} {New York},\ \bibinfo {year} {2009})\BibitemShut {NoStop}%
\bibitem [{\citenamefont {van Kampen}(2007)}]{Kampen2007}%
  \BibitemOpen
  \bibfield  {author} {\bibinfo {author} {\bibfnamefont {N.~G.}\ \bibnamefont
  {van Kampen}},\ }\href@noop {} {\emph {\bibinfo {title} {{Stochastic
  Processes in Physics and Chemistry}}}},\ \bibinfo {edition} {3rd}\ ed.\
  (\bibinfo  {publisher} {Elsevier Science},\ \bibinfo {address} {Amsterdam},\
  \bibinfo {year} {2007})\BibitemShut {NoStop}%
\end{thebibliography}%

%%%%%%%%%%%%%%%%%%%%%%%%%%%%%%%%%%%%%%%%%%%%%%%
%\pagebreak
\clearpage
\widetext

\begin{center}
\textbf{\large Supplemental Materials}
\end{center}
\setcounter{equation}{0}
\setcounter{figure}{0}
\setcounter{table}{0}
\setcounter{page}{1}

\subsection{Linear response of stochastic reactive systems}
\subsubsection{Linear Fokker-Planck equation and its stationary distribution}
In the main text, we encounter multiple times the linear stochastic differential equation (SDE) of the form
\begin{smequation} \label{slinsde1}
	\frac{d \vec y}{dt} = \mx A \vec y + \vec \eta(t),
\end{smequation}%
where $\mx A$ is independent of $\vec y$ and $\vec \eta$ are Gaussian white noises with zero mean and correlator
\begin{smequation}
	\langle \vec \eta(t)\, \vec \eta^T(t') \rangle = \mx B \delta(t-t').
\end{smequation}%
The noise matrix $\mx B$ is symmetric (\textit{i.e} $\mx B^T = \mx B$) and also supposed independent of $\vec y$. Equation~\eqref{slinsde1} is tantamount to the Fokker-Planck equation for the probability density $P(\vec y, t)$~\cite{Gardiner2009}:
\begin{smequation} \label{fpe1}
	\frac{\partial P(\vec y, t)}{\partial t} = -  \sum_{i,j} A_{ij} \frac{\partial}{\partial y_i} (y_j P) + \inv 2  \sum_{i,j} \frac{\partial^2}{\partial y_i \partial y_j} (B_{ij} P).
\end{smequation}%
As shown in (\textit{e.g.})~\cite{Kampen2007}, the stationary distribution is Gaussian and takes the form
\begin{smequation} \label{fpe_stat}
	P_s(\vec y) =  \frac{1}{\sqrt {\det(2 \pi \mx \Xi)}} \exp \left( -\frac{1}{2} \vec y^T\, \mx\Xi^{-1}\, \vec y \right),
\end{smequation}%
where the covariance matrix $\mx \Xi$ is symmetric and given by the Sylvester's equation,
\begin{smequation} \label{sylv}
\mx A \mx \Xi + \mx \Xi \mx A^T + \mx B = 0.
\end{smequation}%
In two dimensions, this equation can be solved~\cite{Gardiner2009} leading to an explicit formula for $\mx \Xi$:
\begin{smequation}\label{Xi2by2}
	\mx \Xi = \frac{\left(\mx A - \id_2\,\text{tr} \mx A\right) \mx B\left(\id_2\,\text{tr} \mx A - \mx A \right)^T - \mx B\,\det \mx A}{2\, \text{tr} \mx A \,\text{det} \mx A}.
\end{smequation}%

%%%%%%%%%%%%%%%%%%%%%%%%%%

\subsubsection{The mean amplification factor $\langle \norm{\vec y}^2 \rangle$}\label{sec2}
We now wish to find an expression for the mean amplification factor, $\langle \norm{\vec y}^2 \rangle$, used in the main text to quantify the linear response of a stochastic reactive system. The norm of $\vec y$ is the Euclidean norm $\norm{\vec y} = \sqrt{\sum_i \left|y_i^2\right|}$. Specifically, we want to compute the integral:
\begin{smequation} \label{norm1}
	\langle \norm{\vec y}^2 \rangle = \int_{\mathbb{R}^D} d\vec y\, P_s(\vec y) \norm{\vec y}^2,
\end{smequation}%
where the distribution $P_s(\vec y)$ is given by Eq.~\eqref{fpe_stat}. Therefore,
\begin{smequation} \label{norm2}
	\langle \norm{\vec y}^2 \rangle = \frac{1}{\sqrt {\det(2 \pi \mx \Xi)}} \int d\vec y \exp \left( -\frac{1}{2} \vec y^T\, \mx\Xi^{-1}\, \vec y \right) \norm{\vec y}^2.
\end{smequation}%
To evaluate this integral, we use the identity
\begin{smequation}\label{eq:integral_id}
	\int \norm{\vec p}^2 e^{-\vec p^{\,T} \mx M \vec p} d\vec p = \inv 2 \Tr\left(\mx M^{-1}\right)\int  e^{-\vec p^{\,T} \mx M \vec p} d\vec p,
\end{smequation}%
with $\mx M = 1/2\mx\Xi^{-1}$, which yields the compact expression:
\begin{smequation} \label{norm3}
	\langle \norm{\vec y}^2 \rangle = \, \Tr\,\mx\Xi%\frac{\text{tr}\,\mx \Xi^{-1}}{\text{det}\,\mx \Xi^{-1}}.
\end{smequation}%

In the following, we assume for convenience that the noise matrix $\mx B$ is a multiple of identity identity matrix $\id$ ($\mx B = \sigma^2 \id$), a choice that can be made without losing in generality. In fact, since $\mx B$ is symmetric, it is diagonalized by an orthogonal matrix which one can use to transform the noises; the resulting diagonal matrix can then be mapped to the identity matrix simply by rescaling the variables $\vec y$. Now, we will write the matrix $\mx \Xi$ in terms of $\mx A$ and what we call the Hermitianizer of $\mx A$, defined as 
\begin{smequation} \label{matG}
	\mx G = - \frac{1}{2}\,\sigma^2\, \mx \Xi^{-1} \mx A^{-1},
\end{smequation}%
which yields a symmetrization of matrix $\mx A$: even though $\mx A$ is not symmetric, $\mx A \ne \mx A^T$, the product $\mx G \mx A = -2^{-1}\sigma^2 \mx \Xi^{-1}$ is a symmetric matrix. Sylvester equation \eqref{sylv} written in terms of $\mx G$ simplifies to
\begin{smequation} \label{eq:G1}
	\inv 2 (\mx G^{-1} + \mx G^{-T}) = \id,
\end{smequation}%
indicating that the hermitian part of $\mx G^{-1}$ is identity. Alternatively, the Hermitianizer of $\mx A$ can be defined as the unique matrix satisfying \eq{G1} whose product with $\mx A$ is Hermitian. Now we can write the mean squared value of the norm $\vec y$ in terms of $\mx A$ and $\mx G$ by substituting Eq.~\eqref{matG} in Eq.~\eqref{norm3}:
\begin{smequation} \label{norm4}
	\langle \norm{\vec y}^2 \rangle = -\inv 2 \sigma^2\, \Tr\left(\mx A^{-1} \mx G^{-1}\right)
\end{smequation}%

When $\mx A$ is a $2\times 2$ matrix, the trace of the inverse can be written as trace over determinant:
\begin{smequation} \label{norm2by2}
	\langle \norm{\vec y}^2 \rangle = -\inv 2 \sigma^2\, \frac{\Tr\left(\mx G \mx A\right)}{\det(\mx G)\det(\mx A)}
\end{smequation}%
$\Tr(\mx G \mx A)$ can be simplified by taking the trace of Eq.~\eqref{matG}
\begin{smequation} \label{trGA}
	\Tr(\mx G\mx A) = -\inv 2\,\sigma^2\, \Tr(\mx \Xi^{-1}).
\end{smequation}%
Also, by multiplying the right-hand side of the Sylvester equation~\eqref{sylv} by $\mx \Xi^{-1}$:
\begin{smequation}
	\mx A + \mx \Xi \mx A^T \mx \Xi^{-1} = - \sigma^2\,\mx \Xi^{-1}.
\end{smequation}%
and taking the trace we have (recalling that $ \Tr (\mx \Xi \mx A^T \mx \Xi^{-1})=\Tr (\mx A^T)= \Tr (\mx A)$):
\begin{smequation} \label{trinvxi}
	\sigma^2\,\Tr(\mx \Xi^{-1}) = - 2\, \Tr (\mx A)
\end{smequation}%
From Eq.~\eqref{trinvxi} and Eq.~\eqref{trGA} it follows that $\Tr(\mx G\mx A) = \Tr(\mx A)$.
which we can use to simply Eq.~\eqref{norm4}:
\begin{smequation} \label{norm5}
	\langle \norm{\vec y}^2 \rangle = -\frac{\,\sigma^2}{2\,\det\mx G} \frac{\text{Tr}\,\mx A}{\text{det}\,\mx A} = -\inv 2\,\sigma^2\, \det\left(\mx G^{-1}\right) \Tr\left(\mx A^{-1}\right).
\end{smequation}%

%%%%%%%%%%%%%%%%%

\subsubsection{Non-normality for a $2\times2$ matrix $\mx A$}
For a $2\times 2$ matrix $\mx A$ given by its elements
\begin{smequation}
	\mx A = \left(\begin{array}{cc}
			a_{11}	& a_{12}\\
			a_{21}	& a_{22}
		\end{array}\right),
\end{smequation}%
we can solve for $\mx \Xi$ from Eq.~\eqref{Xi2by2} and substitute in Eq.~\eqref{matG} to find the matrix $\mx G$ in terms of matrix elements of $\mx A$:
\begin{smequation}
	\mx G = \left(\begin{array}{cc}
		 	\frac{(a_{11}+a_{22})^2}{(a_{12}-a_{21})^2+(a_{11}+a_{22})^2} & -\frac{(a_{12}-a_{21}) (a_{11}+a_{22})}{(a_{12}-a_{21})^2+(a_{11}+a_{22})^2} \\
			\frac{(a_{12}-a_{21}) (a_{11}+a_{22})}{(a_{12}-a_{21})^2+(a_{11}+a_{22})^2} & \frac{(a_{11}+a_{22})^2}{(a_{12}-a_{21})^2+(a_{11}+a_{22})^2} \\
		\end{array}\right).
\end{smequation}%
The non-normality index $\mathcal{H}$ is given by the inverse of the determinant of $\mx G$:
\begin{smequation}\label{simple_index}
	\mathcal{H}(\mx A) = \det\left(\mx G^{-1}\right) = 1 + \frac{(a_{12}-a_{21})^2}{(a_{11}+a_{22})^2}.
\end{smequation}%

If the eigenvalues of $\mx A$ are real, we can rewrite this expression in terms of the eigenvalues and the angle between the eigenvectors of $\mx A$. Let $\Delta>0$ be the discriminant of the characteristic polynomial of $\mx A$:
\begin{smequation}
	\Delta = (a_{11}-a_{22})^2 + 4\, a_{12} \, a_{21}.
\end{smequation}%
If $\lambda_1$ and $\lambda_2$ are the two eigenvalues of $\mx A$, and $\vec v_1$ and $\vec v_2$ are the two eigenvectors, we have
\begin{smequation}
\begin{split}
	&\left(\lambda_1 + \lambda_2\right)^2 = (a_{11}+a_{22})^2, \qquad \left(\lambda_1 - \lambda_2\right)^2 = \Delta, \\
	&\cos^2(\theta) = \left(\frac{\vec v_1 \cdot \vec v_2}{\norm{\vec v_1}  \norm{\vec v_2}}\right)^2, \qquad \cot^2(\theta) = \frac{\cos^2(\theta)}{1-\cos^2(\theta)} = \frac{ (a_{11}-a_{22})^2}{\Delta}.
\end{split}
\end{smequation}%
Now it is clear that 
\begin{smequation}
	\mathcal{H}(\mx A) = 1 + \cot^2(\theta)\left( \frac{ \lambda_1 - \lambda_2}{ \lambda_1 + \lambda_2}\right)^2.
\end{smequation}%

%%%%%%%%%%%%%%%%%%%%%%%%%%

\subsubsection{Linear stochastic differential equations with complex variables}
Consider a similar set of SDEs of the the form
\begin{smequation} \label{eq:c_sde}
	\frac{d \vec y}{dt} = \mx A \vec y + \vec \eta(t),
\end{smequation}%
where now $\vec y$ and $\vec \eta$ are vectors with complex variables, and $\vec \eta$ is a Gaussian white noise with zero mean and correlator
\begin{smequation}
\begin{split}
	&\langle \vec \eta(t)\, \vec \eta\,^\dagger(t') \rangle = \mx B \delta(t-t'),\\
	&\langle \vec \eta(t)\, \vec \eta\,^T(t') \rangle = 0.
\end{split}
\end{smequation}%
where the $^\dagger$ symbol represents the transpose conjugate. The analysis in the previous section can be generalized by evaluating the expected value of $\vec y(t) \vec y\,^\dagger(\tau)$ and $\vec y(t) \vec y\,^T(\tau)$ at steady state for $t = \tau$ to obtain the following relationships for the {\it covariance} and {\it relation} matrices
\begin{smequation}
\begin{split}
	&\mx A\left\langle \vec y \vec y\,^\dagger \right\rangle + \left\langle \vec y \vec y\,^\dagger \right\rangle  \mx A^\dagger + \mx B = 0\\
	&\mx A\left\langle \vec y \vec y\,^T \right\rangle + \left\langle \vec y \vec y\,^T \right\rangle  \mx A^T = 0
\end{split}
\end{smequation}%
The first equation is the analogue of equation of Sylvester Eq.~\eqref{sylv} for the Hermitian covariance matrix $\mx \Xi = \left\langle \vec y \vec y\,^\dagger \right\rangle$, while the second equation implies that the symmetric relation matrix $\mx C = \left\langle \vec y \vec y\,^T \right\rangle$ is equal to zero. Therefore, at steady state, $\vec y$ obeys a circularly symmetric complex Gaussian distribution of the form 
\begin{smequation} 
	P_s(\vec y) =  \frac{1}{\det(2 \pi\mx \Xi)} \exp \left( -\frac{1}{2} \vec y\,^\dagger\, \mx\Xi^{-1}\, \vec y \right).
\end{smequation}%
Notice the different normalization factor compared to Eq~\eqref{fpe_stat}, as it is normalized over $\mathbb{C}^D$ instead of $\mathbb{R}^D$.

To compute the mean square value of the norm of $\vec y$, we can follow similar analysis to that of section \ref{sec2}. Here, we highlight the differences. The mean square norm is define as
\begin{smequation} \label{eq:cnorm1}
	\langle \norm{\vec y}^2 \rangle = \int_{\mathbb{C}^D} d\vec y\, P_s(\vec y) \norm{\vec y}^2,
\end{smequation}%
with the norm $\norm{\vec y} = \sqrt{\vec y\,^\dagger \vec y}$. The complex version of \eq{integral_id} can be evaluated by diagonalizing the matrix $\mx M$ and write the integral on a $2D$-dimensional real space. The result is given by
\begin{smequation}\label{eq:cintegral_id}
	\int_{\mathbb{C}^D} \norm{\vec p}^2 e^{-\vec p^{\,\dagger} \mx M \vec p} d\vec p = \Tr\left(\mx M^{-1}\right)\int_{\mathbb{C}^D}  e^{-\vec p^{\,\dagger} \mx M \vec p} d\vec p\,,
\end{smequation}%
where the factor $1/2$ is canceled by the fact that each eigenvalue of $\mx M^{-1}$ should be counted twice in the $2D$-dimensional space, once for the real part and once for the imaginary part. As a result, there will be an extra factor $2$ in Eq.~\eqref{norm3}, Eq.~\eqref{norm4}, and Eq.~\eqref{norm5}. In particular ,
\begin{smequation} \label{eq:cnorm2}
	\langle \norm{\vec y}^2 \rangle = - \sigma^2\, \Tr\left(\mx A^{-1} \mx G^{-1}\right)
\end{smequation}%

%%%%%%%%%%%%%%%%%
\subsection{Analysis of model by Ridolfi et al.}
\subsubsection{From individual level model to SDEs}
In this section we derive a the stochastic extension of the model by Ridolfi et al.~\cite{ridolfi2011transient} by expanding the master equation corresponding to the individual level model defined by the following set of reactions
\begin{smequation}\label{rx:smain}
\begin{split}
	&2U_i+V_i\xrightarrow{a}3U_i+V_i,\hspace{0.05\textwidth}
	V_i\xrightarrow{b}2V_i,\\
	&U_i\xrightarrow{e}\varnothing,\hspace{0.05\textwidth}
	2V_i+2U_i\xrightarrow{c}V_i+2U_i,
\end{split}
\end{smequation}%
where $U_i$ and $V_i$ are the species $U$ and $V$ in the site $i$, and the diffusion reactions
\begin{smequation}\label{rx:smain}
	U_i\xrightarrow{\delta_u}U_j,\hspace{0.05\textwidth}
	V_i\xrightarrow{\delta_v}V_j,\hspace{0.05\textwidth} j\in\langle i\rangle
\end{smequation}%
where $\langle i\rangle$ is the set of sites neighboring $i$, $\delta_u=D_U/\Omega^{2/D}$, $\delta_v=D_V/\Omega^{2/D}$, $D_U$ and $D_V$ are the diffusion constants, and $\Omega$ is the volume of each site. The state of the system is specified by the concentration vectors $\vec q_i\equiv (u_i, v_i) \equiv (U_i, V_i)/\Omega$.

Each reaction of reaction scheme \rx{smain} takes the system from a state $\{\vec q_i\}$ to $\{\vec q_i\!\rq{}\}$ with probability per unit time $T(\{\vec q_i\!\rq{}\}|\{\vec q_i\})$. These transition rates are given from the law of mass action:
\begin{smequation}
\begin{split}
	&T\left(\left. \vec q_i +\vec s_1\right|\vec q_i\right) = \Omega a  u_i^2 v_i, \hspace{0.03\textwidth}
	T\left(\left. \vec q_i + \vec s_2\right|\vec q_i\right) = \Omega b v_i,\\
	&T\left(\left. \vec q_i - \vec s_1\right|\vec q_i\right) = \Omega e  u_i,\hspace{0.03\textwidth}
	T\left(\left. \vec q_i - \vec s_2\right|\vec q_i\right) = \Omega c  u_i^2 v_i^2,\\
\end{split}
\end{smequation}%
and for every $j\in\langle i\rangle$
\begin{smequation}
\begin{split}
	&T\left(\left. \vec q_i -\vec s_1,\vec q_j + \vec s_1\right|\vec q_i,\vec q_j\right) = \Omega \delta_u u_i,\\
	&T\left(\left. \vec q_i -\vec s_2,\vec q_j + \vec s_2\right|\vec q_i,\vec q_j\right) = \Omega \delta_v v_i,
\end{split}
\end{smequation}%
where
\begin{smequation}
	\vec s_1 = \Omega^{-1} \left(\begin{array}{c}
		1\\0
	\end{array}\right), \hspace{0.05\textwidth}
	\vec s_2 = \Omega^{-1} \left(\begin{array}{c}
		0\\1
	\end{array}\right).
\end{smequation}%
The master equation for the time evolution of the probability of finding the system at a state $\{\vec q_i\}$, $P(\{\vec q_i\},t)$ can be written as
\begin{smequation}\label{eq:smaster}
	\der{P(\{\vec q_i\},t)}{t} = \sum_{\{\vec q_i\!\rq{}\}} \left(T(\{\vec q_i\}|\{\vec q_i\!\rq{}\})-T(\{\vec q_i\!\rq{}\}|\{\vec q_i\})\right)
\end{smequation}%
Following \cite{mckane2014stochastic}, we can expand the right hand side of \eq{smaster} to second order in $\Omega^{-1}$ obtaining a Fokker-Planck equation corresponding the following set of stochastic differential equations
\begin{smequation}
\begin{split}\label{eq:slangevin}
	&\der{u_i}{t} = u_i(a u_i v_i-e)+\delta_u\sum_{j\in\langle i\rangle}(u_j-u_i)+\xi_i(t),\\
	&\der{v_i}{t} = v_i(b-c u_i^2 v_i)+\delta_v\sum_{j\in\langle i\rangle}(v_j-v_i)+\eta_i(t),\\
\end{split}
\end{smequation}%
where $\xi_i$\rq{}s and $\eta_i$\rq{}s are zero mean Gaussian noise with correlations
\begin{smequation}\label{eq:snoise1}
\begin{split}
	\langle\xi_i(t)\xi_j(t\rq{})\rangle &= \frac{\delta(t-t\rq{})}{\Omega}\Bigg(\bigg( u_i(a u_i v_i+e)+\delta_u\sum_{k\in\langle i\rangle}(u_i+u_k)\bigg)\delta_{i,j}-\delta_u(u_i+u_j)\Chi_{\langle i\rangle}(j)\Bigg)\\
	\langle\eta_i(t)\eta_j(t\rq{})\rangle &= \frac{\delta(t-t\rq{})}{\Omega}\Bigg(\bigg( v_i(b+c u_i^2 v_i)+\delta_v\sum_{k\in\langle i\rangle}(v_i+v_k)\bigg)\delta_{i,j}-\delta_v(v_i+v_j)\Chi_{\langle i\rangle}(j)\Bigg)
\end{split}
\end{smequation}%
and the characteristic function, $\Chi_{\langle i\rangle}$, of $\langle i\rangle$ is defined as
\begin{smequation}
	\Chi_{\langle i\rangle}(j) = \begin{cases}
		1	&	j\in \langle i\rangle\\
		0	&	j\notin \langle i\rangle
	\end{cases}\;.
\end{smequation}%
By defining $\vec f(\vec q) \equiv (f,\; g) \equiv ( u(a u v-e), \;v(b-c u^2 v))$, $\vec \xi_i \equiv (\xi_i,\; \eta_i)$, $\mx \delta \equiv \mx{diag}(\delta_u,\;\delta_v)$, and $\left(\Delta \vec q\right)_i \equiv \sum_{j\in\langle i\rangle}(\vec q_j-\vec q_i)$, \eq{slangevin} can be written in the simple form
\begin{smequation}\label{eq:snonlinear}
	\der{\vec q_i}{t} = \vec f(\vec q_i)+\mx\delta\left(\Delta\vec q\right)_i+\vec\xi_i(t).
\end{smequation}%
\Eq{snonlinear} is the discrete space version of \eq{nonlinear} of the main text. Continuous limit can be taken at any point in the following analysis to recover the continuous space stochastic partial differential equations of type analyzed in the main text. We continue with the discrete version where the analytic results can be more readily compared to the simulation.

The deterministic part of our model has a fixed point $\vec q\,^* \equiv (u^*,\; v^*) = (ba/ce,\; e^2c/a^2b)$, obtained by setting $\vec f(\vec q)$ equal to zero. We can linearize \eq{snonlinear} around the fixed point $\vec q\,^*$, by defining $\vec p_i \equiv \big((u_i-u^*)/\sqrt{2u^*e},\,(v_i-v^*)/\sqrt{2v^*b}\big)$  which are the rescaled deviations of $\vec q_i$ from $\vec q\,^*$,
\begin{smequation}\label{eq:slinear}
	\der{\vec p_i}{t} = \mx J\vec p_i+\mx \delta(\Delta\vec p)_i + \vec\xi_i(t),
\end{smequation}%
where the linear stability operator $\mx J$ is defined as the Jacobian of the transformed function $f$ at the fixed point $\vec p = 0$ is given by
\begin{smequation}\label{eq:sjacobian}
	\mx J = \left(\begin{array}{cc}
		e 	&	\frac{b^\frac{3}{2}a^\frac{3}{2}}{ce}\\%\frac{b^2a^3}{c^2e^2}\\
		-\frac{2e^2 c}{a^\frac{3}{2} b^{\inv 2}}	& -b%-\frac{2e^3c^2}{a^3b}	&	-b
	\end{array}\right)
\end{smequation}%
Evaluating \eq{snoise1} at $\vec q\,^*$
\begin{smequation}\label{eq:snoise2}
\begin{split}
	\langle\xi_i(t)\xi_j(t\rq{})\rangle &= \frac{\delta(t-t\rq{})}{\Omega}\Big(\big( 1+\delta_u n/e\big)\delta_{i,j}-\delta_u\Chi_{\langle i\rangle}(j)\Big),\\
	\langle\eta_i(t)\eta_j(t\rq{})\rangle &= \frac{\delta(t-t\rq{})}{\Omega}\Big(\big( 1+\delta_v n/b\big)\delta_{i,j}-\delta_v\Chi_{\langle i\rangle}(j)\Big),
\end{split}
\end{smequation}%
where $n \equiv \left|\langle i\rangle\right|$ is the number of neighbors of each site. Note that for $b>e$, both of the eigenvalues of $\mx J$ have negative real parts, making $\vec q\,^*$ an attractor of the dynamics in the absence of the diffusion.

To examine the spatial stability of $\vec q\,^*$, we need to diagonalize the discrete Laplacian operator $\Delta$, by defining the discrete Fourier transform of a sequence $\{s_{\vec n}\}$ as
\begin{smequation}
\tilde s_{\vec k} \equiv \left(\mathcal F[\{s_{\vec n}\}]\right)_{\vec k} \equiv \inv{\sqrt{N^D}}\sum_{\vec n} \e^{-2\pi\vec k.\vec n/N}s_{\vec n}.
\end{smequation}%
We drop the tildes on the Fourier variable with the convention that the variables with index $k$ are Fourier variables. \Eq{slinear} under this transformation becomes
\begin{smequation}\label{eq:sdiagonal}
	\der{\vec p_{\vec k}}{t} =\mx K\vec p_{\vec k} + \vec\xi_{\vec k}(t), \hspace{0.03\textwidth} \mx K =  \mx J+\Delta(\vec k) \mx \delta,
\end{smequation}%
where $\Delta(\vec k)$ is the discrete Fourier transform of the discrete Laplacian operator given by
\begin{smequation}
	\Delta(\vec k ) \equiv -2\sum_{l = 1}^D\big(1-\cos(2\pi k_l/N)\big)
\end{smequation}%
and
\begin{smequation}\label{eq:snoise2}
\begin{split}
	\langle\xi_{\vec k}(t)\xi_{\vec k\rq{}}^*(t\rq{})\rangle = \Omega^{-1}\left(1-e^{-1}\delta_u\Delta(\vec k)\right)\delta_{\vec k,\vec k\rq{}}\delta(t-t\rq{}),\\
	\langle\eta_{\vec k}(t)\eta_{\vec k\rq{}}^*(t\rq{})\rangle =\Omega^{-1}\left(1-b^{-1}\delta_v\Delta(\vec k)\right)\delta_{\vec k,\vec k\rq{}}\delta(t-t\rq{}).\\
\end{split}
\end{smequation}%

For the regime that we observe stochastic patterns, the contribution of the diffusion process in the amplitude of the noise in \eq{snoise2} is very small and will be neglected for simplicity. This approximation is not necessary, since there is always a change of variables that simplifies the correlation matrix to a multiple of the identity matrix (this is the reason for the rescaling in the definition of $\vec p$). With this approximation 
\begin{smequation}\label{eq:snoise3}
	\left\langle\vec \xi_{\vec k}(t)\vec \xi_{\vec k\rq{}}^{\,\dagger}(t\rq{})\right\rangle = \Omega^{-1}\delta_{\vec k,\vec k\rq{}}\delta(t-t\rq{})\,\id
\end{smequation}%
where $\vec \xi_{\vec k\rq{}}^{\,\dagger}$ is the conjugate transpose of $\vec \xi_{\vec k\rq{}}$, and $\id$ is the $2\times 2$ identity matrix.

%%%%%%%%%%%%%%%%%%%%

\subsubsection{Phase diagram of pattern formation}
The pattern forming behavior of the model defined by \rx{smain} can be understood by analyzing the eigenvalues of $\mx K$ as a function of $\vec k$. Matrix $\mx K$ can be written in elements from \eq{sdiagonal} and \eq{sjacobian}:
\begin{smequation}
	\mx K = \left(\begin{array}{cc}
		e + \Delta(\vec k) \delta_u	&	\frac{b^\frac{3}{2}a^\frac{3}{2}}{ce}\\
		-\frac{2e^2 c}{a^\frac{3}{2} b^{\inv 2}}	& -b + \Delta(\vec k) \delta_v
	\end{array}\right)
\end{smequation}%
As it will become clear, most of the properties of the system depend on the following three parameters
\begin{smequation}
	\rho = \frac{b}{e},\qquad \nu = \frac{e\, c}{a^{\frac{3}{2}} b^{\inv 2}}, \qquad r = \frac{\delta_v}{\delta_u} = \frac{D_V}{D_U}
\end{smequation}%
in the following analysis, we will write various expression in terms of these parameters, wherever we can. We start with $\mx K$
\begin{smequation}
	\mx K = \left(\begin{array}{cc}
		e + \Delta(\vec k) \delta_u	&	b/\nu\\
		-2 e \,\nu	& -b + \Delta(\vec k) \delta_v
	\end{array}\right)
\end{smequation}%
The largest eigenvalue of $\mx K$ is given by
\begin{smequation}\label{eq:slambda}
	\lambda(\vec k) = \frac{1}{2} \left(\sqrt{b^2-2 b \Delta(\vec k) (\delta_v-\delta_u)-6 b e+\left(e-\Delta(\vec k) (\delta_v-\delta_u)\right)^2}-b+\Delta(\vec k) (\delta_v+\delta_u)-e\right).
\end{smequation}%
Notice that the eigenvalues of $\mx K$ are independent of $\nu$. For small $\vec k$, $\Delta(\vec k)$ is a monotonically decreasing function of $\vec k$ (proportional to $-k^2$). We define $y = -\Delta(\vec k)$. To determine if $\lambda$ monotonically decays or if it has a maximum at some $\vec k_0 \neq 0$, we can differentiate $\lambda$ with respect to $y$ and see if it has a positive root. The largest root of $\der{\lambda}{y}$ is given by
\begin{smequation}\label{eq:y0}
	y_0 = -\Delta(\vec k_0) = \frac{ (r+1) \sqrt{2\, b\, e\, r}-b\, r-e\, r}{\delta_u \,(r-1)\, r}.
\end{smequation}%
For $y_0$ to be greater than zero we need
\begin{smequation}
	\rho < \frac{\left(1+ r+r^2+(r+1) \sqrt{r^2+1}\right)}{r}.
\end{smequation}%
We can find the condition on the ratio of the diffusion constants by inverting this inequality:
\begin{smequation}
	r > \frac{1-2\, \rho + \rho^2+(1+ \rho) \sqrt{1 + \rho\, (\rho-6)}}{4 \,\rho} = f_1(\rho).
\end{smequation}%

The condition for formation of stochastic pattern is $\lambda(\vec k_0) > \Re(\lambda(0))$. We can find $\lambda(\vec k_0)$ and $\lambda(0)$ by substituting $y_0 = y(\vec k_0)$ from \eq{y0}  and $y(0) = 0$ in \eq{slambda}:
\begin{smequation}
	\lambda(\vec k_0) = \frac{b+e\, r- \sqrt{8\,b\, e\, r}}{r-1}, \qquad \lambda(0) = \frac{1}{2} \left(\sqrt{b^2-6\, b\, e+e^2}-b+e\right).
\end{smequation}%
Then, $\lambda(\vec k_0) > \Re(\lambda(0))$ simplifies to 
\begin{smequation}
	r > \frac{- 1 + 14\, \rho  - \rho^2+4 \sqrt{-2\,\rho\, (1 + \rho\, (\rho-6)))}}{(1 + \rho)^2} = f_2(\rho).
\end{smequation}%
Condition for deterministic Turing pattern is a lot simpler; we just need $\lambda(\vec k_0) > 0$ which simplifies to
\begin{smequation}
	r > \left(3+2 \sqrt{2}\right) \rho   = f_3(\rho).
\end{smequation}%

When $r$ is greater than $f_1(\rho)$ and $f_2(\rho)$ but less than $f_3(\rho)$, the system exhibits stochastic patterns (blue region in Fig. 3 of the main text), while we observe the deterministic patterns when $r$ is greater than $f_3$ (orange region of Fig. 3 of the main text).

%%%%%%%%%%%%%%%%%%%%

\subsubsection{Non-normality of the model}
The amplification of our stochastic patterns depend on the non-normality index of $\mx K_0 = \mx K(\vec k_0)$ given by
\begin{smequation}
	\mx K_0 = \left(\begin{array}{cc}
		e - y_0\, \delta_u	&	b/\nu\\
		-2\, e \,\nu	& -b - y_0\, \delta_v
	\end{array}\right),
\end{smequation}%
where $y_0 = -\Delta(\vec k_0)$. We use Eq.~\eqref{simple_index} to calculate the non-normality index of $\mx K_0$:
\begin{smequation}
	\mathcal{H}(\mx K_0) = 1 + \left(\frac{b+2 e \nu ^2}{\nu (b-e+y_0\, (\delta_u+\delta_v))}\right)^2.
\end{smequation}%
We substitute $y_0$ from \eq{y0} and rewrite the resulting expression in terms of $\rho$, $r$, and $\nu$:
\begin{smequation}
	\mathcal{H}(\mx K_0) = 1+\left(\frac{2 \nu ^2+\rho }{\nu \left(\rho - 1 +\frac{(r+1) \left(-\rho  r+ (r+1) \sqrt{2\rho\,  r}-r\right)}{(r-1) r}\right)}\right)^2
\end{smequation}%
Since the eigenvalues of $\mx K$ do not depend on $\nu$, one can change $\mathcal{H}(\mx K_0)$ by changing $\nu$ without moving the system in its phase diagram (see Fig. 3 of the main text). This can be done by changing the ratio of $a/c^{2/3}$ without affecting $\rho$.

\end{document}